\documentclass[lettersize,journal]{IEEEtran}

\usepackage{algorithmic}
\usepackage{algorithm}
\usepackage{array}
\usepackage[caption=false,font=normalsize,labelfont=sf,textfont=sf]{subfig}
\usepackage{amsthm}
\usepackage{cite}
\usepackage{amsmath,amssymb,amsfonts}
\usepackage{graphics}
\usepackage{graphicx}
\usepackage{textcomp}
\usepackage{float}
\usepackage{epsfig} 
\usepackage{cite}
\usepackage{textcomp}
\usepackage{stfloats}
\usepackage{url}
\usepackage{verbatim}
\usepackage{graphicx}

\newtheorem{theorem}{Theorem}[section]
\newtheorem{lemma}[theorem]{Lemma}

\begin{document}

\title{Position Control of Single Link Flexible Manipulator: A Functional Observer Based Sliding Mode Approach}

\author{Atul Sharma$^{a}$(\textit{Student Member, IEEE}),
S Janardhanan$^{b}$ (\textit{Senior Member, IEEE})
\thanks{$^{a}$Research Scholar, Department of Electrical Engineering, Indian Institute of Technology, Delhi; \texttt{atul.sharma@ee.iitd.ac.in}}
\thanks{$^{b}$Professor, Department of Electrical Engineering, Indian Institute of Technology, Delhi; \texttt{janas@ee.iitd.ac.in}}
}

\markboth{Journal of \LaTeX\ Class Files,~Vol.~14, No.~8, August~2021}%
{Shell \MakeLowercase{\textit{et al.}}: A Sample Article Using IEEEtran.cls for IEEE Journals}


\maketitle

\begin{abstract}
This paper proposes a functional observer-based sliding mode control technique for position control of a single-link flexible manipulator. The proposed method considers the unmodelled system dynamics as uncertainty and aims to achieve accurate position control. The functional observer is used to directly estimate the sliding mode control design components and a sliding mode controller to generate the control signal, which guarantees the system's robustness and stability. The proposed control scheme is validated using numerical simulations.
\end{abstract}
\textbf{Note to Practitioners:}


\section{Introduction}\label{sec1}
In recent years, the robotic manipulators have been explored for a wide range of applications including industrial production \cite{yuan2007research}, hostile environments (nuclear sites, deep sea, etc.) \cite{kress1997waste}, space exploration \cite{stieber1999vision}, health care equipment \cite{lomanto2015flexible}, building construction \cite{warszawski1985robotics}, etc. It is required that the robotic manipulators should provide faster, cost-effective and accurate operation \cite{xiao2016tracking,alandoli2020critical}. The rigid robot manipulators are made up of rigid links which makes them bulkier \cite{sun2018fuzzy}. The industries need an upgrade to the existing classical robots in order to reduce construction costs, minimize energy consumption brought on by big actuator sizes, and increased production. 
\par So in applications where there is a requirement that the weight-to-volume ratio of a manipulator be low, inevitably, the manipulator tends to be flexible. There are applications where large and lighter manipulators are required \cite{mejerbi2018dynamic}, and as a consequence of the larger and lighter arms flexibility comes into the picture. Also, as the payload-to-weight ratio increases, the tendency of flexible modes to get excited increases. The flexibility in the manipulator is modelled as the link deformation \cite{tavasoli2018dynamic}. Hence, analysis of such systems can not be performed as rigid manipulators. If we consider the flexibility, then the system formed will be of infinite dimension, i.e., the dynamic model of the flexible link robot manipulator is described as a distributed parameter system. This makes the dynamics of a flexible link robot manipulator depend on both space and time. Hence, the mathematical analysis of a flexible link manipulator would involve partial differential equations (PDE) rather than the ordinary differential equation (ODE). From a control viewpoint, finding the direct analytical solution to PDEs may not always be possible, and the solutions obtained may not always be realizable. So, we need to approximate the PDE-based mathematical model of the flexible link manipulator to an ODE-based model. There are various approximation methods available in the literature \cite{awrejcewicz2021review} including the finite element method (FEM) \cite{sunada1981application, chedmail1991modelling}, assumed mode method (AMM) \cite{de1991recursive, ata2012dynamic, loudini2013modelling, yang2018dynamic} etc. In this paper, the assumed mode method is chosen over the finite element method \cite{theodore1995comparison}. This is because the finite element method works on the discretization approach; hence, for a lower-order dynamic model, it may not capture the effect of all the potential flexible modes of vibration. 
\par The study of a single-link flexible robot manipulator was the starting point for flexible robot research. There are various methods of modelling available in the literature, Lumped parameter approach \cite{konno1995vibration}, Euler-Bernoulli beam theory \cite{bauchau2009euler}, Hamilton's principle \cite{najafi2017non}, Lagrangian dynamics \cite{book1975feedback}, Newton-Euler-FEM method \cite{naganathan1986non, scaglioni2017closed}, Finite Element Method (FEM) \cite{sunada1981application, chedmail1991modelling}, assumed-modes method \cite{de1991recursive, ata2012dynamic, loudini2013modelling, yang2018dynamic} etc. The most popular approach for building the mathematical model of flexible manipulators is Lagrangian dynamics because the equations of motion are formulated using kinetic and potential energies, which are scalars. Therefore, the equations of motion are derived from a single scalar known as Lagrangian.
\par Due to the flexibility of the link, the tip position of a flexible link robot manipulator depends on both the joint angle and the link deformation variable. Even a small link deformation has a very significant impact on the tip position. Therefore, to perform the specific operation using the flexible link manipulator, a control input must be designed to drive the tip to the desired trajectory. However, due to the deformation in the link, the existing control algorithms are insufficient to efficiently control the flexible link manipulator \cite{alandoli2020critical,akyuz2011fuzzy}.
\par The most desirable characteristics of a control system are a simple design, fast response, and robustness to uncertainties and disturbances. The dynamic model of a flexible link manipulator has inherent, unmodelled uncertainties. Therefore, a robust control design is preferable for such a system. Sliding mode control (SMC) is one of the most used control schemes for uncertain nonlinear systems in order to provide robustness and faster system response \cite{shtessel2014sliding, liu2012advanced}. The SMC scheme is a model-based feedback control technique. This paper uses the state feedback sliding mode control design because of its simplistic design. Therefore, it is required that the system states needed in the control input be available for feedback control design. But in a flexible link manipulator, all the states can never be available for the feedback design. Hence, traditional SMC can not fit such a system well. To overcome this issue, an observer is to be designed that estimates the unmeasurable system state by utilising the knowledge of input and output. However, typically, a linear feedback control law needs estimation of some linear function of states of the form $Kx(t)$. The estimation of the linear function of the state vector can be done using a minimal-order observer. Therefore, a functional observer is designed in this paper to estimate the linear function of the state vector required in the SMC design.
\par A functional observer is a type of model-based control system that uses system outputs to estimate the function of the linear combination of states required in the control input. Previous works have demonstrated several techniques to design functional observers for linear time-invariant (LTI) systems \cite{murdoch1973observer, fahmy1989observers, darouach2000existence, darouach2022functional}. 
\par A functional observer estimates the linear functions of states, which are then used in the sliding mode controller. This composite control strategy is the functional observer-based sliding mode control (FO-SMC) scheme. It has been demonstrated that FO-SMC works well for controlling the position of flexible link manipulators.
\par In this paper, we proposed FO-SMC to control the position of a single-link flexible manipulator. Numerical simulations are utilized to verify the proposed control scheme. The results illustrate that the proposed FO-SMC technique can precisely and successfully control the position of the single-link flexible manipulator.
\par These are some of the main contributions of this paper:
\begin{itemize}
    \item To control the position of the single-link flexible manipulator, a novel composite control approach based on FO-SMC has been proposed. 
    \item Numerical simulation is used to validate the proposed control scheme.
    \item The demonstration of the efficacy of the proposed FO-SMC scheme in controlling the position of a single-link flexible manipulator.
\end{itemize}
The rest of the paper is structured as follows: In Section \ref{sec:modelling}, the dynamic model of the single-link flexible manipulator is presented. The FO-SMC approach for position control is proposed in Section \ref{sec:composite_control}. In Section \ref{sec:simulations}, a discussion on simulation results is presented. Finally, the conclusion is presented in section \ref{sec:conclusion}.
\section{Modelling of Flexible Link Manipulator}
\label{sec:modelling}
Considering a single-link flexible manipulator as shown in figure \ref{fig:Schematic_SFLM} with length $l(m)$, mass is uniformly distributed across the length with linear mass density $\rho(kg/m)$. The following assumptions are considered for modelling the single-link flexible manipulator:
\begin{figure}[h!]
\centering
\includegraphics[width=3.5in,height=2in]{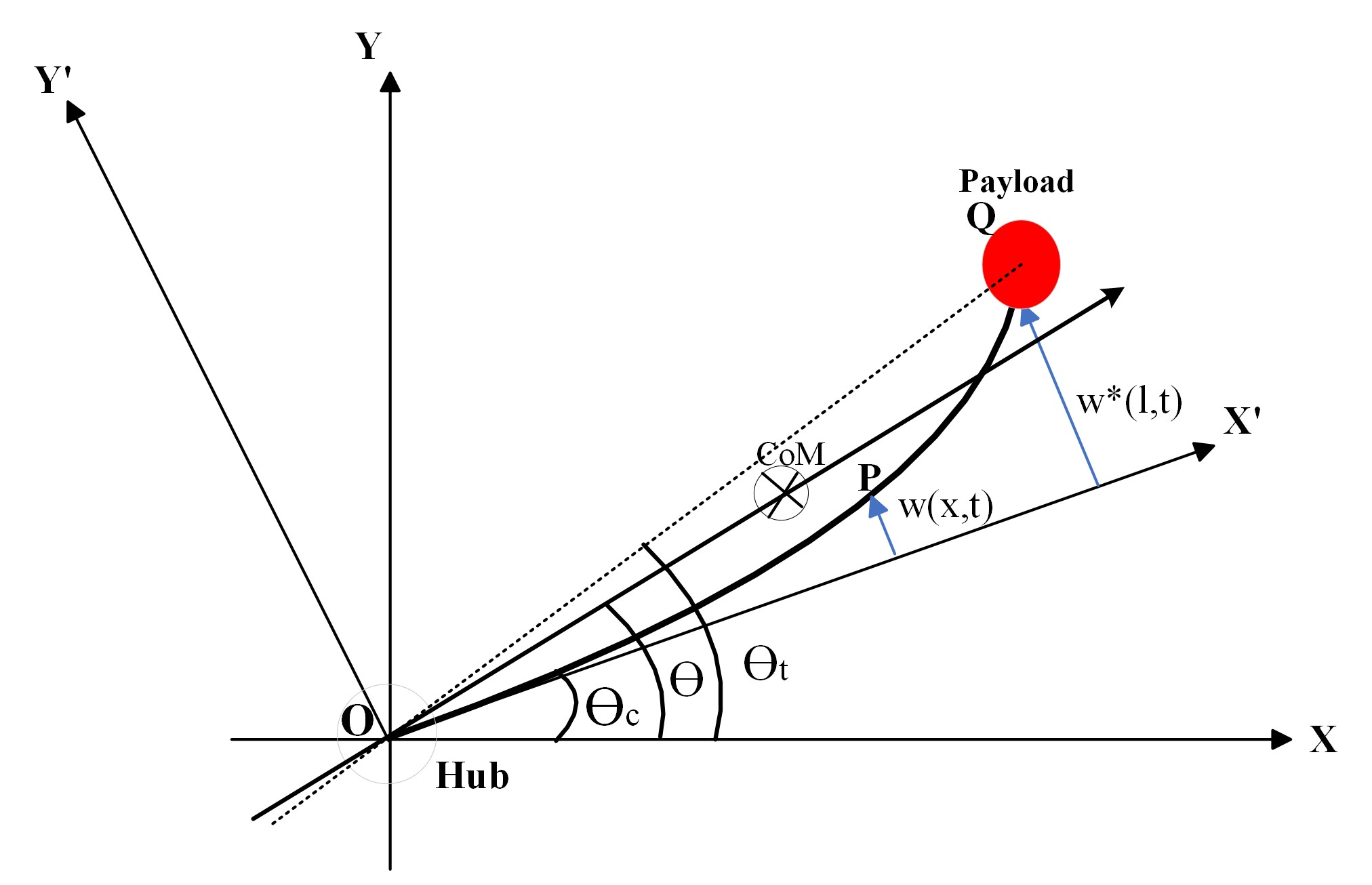}
\caption{Single-Link Flexible Manipulator}
\label{fig:Schematic_SFLM}
\end{figure}
\par \textbf{Assumptions:}
\begin{enumerate}
    \item Mass is uniformly distributed across the length of the link.
    \item Link undergoes only small deformation of pure bending (No torsion and Compression).
    \item Bending forces due to gravity and nonlinear deformations are also negligible.
\end{enumerate}
The flexible link under consideration is modelled as an Euler-Bernoulli beam with $E$ Young's modulus and an $I$ cross-sectional moment of inertia. The electrical motor is connected at the base with inertia $J_0 (kg-m^2)$ provides torque ($\tau$ N-m) to the manipulator, and the payload carried by the manipulator has mass $m_p(kg)$ and inertia $J_p(kg-m^2)$.
\\Using Hamilton's principle and the calculus of variation, it is shown that angle $\theta(t)$ and deformation $w(x,t)$ satisfy the following partial differential equations \cite{de2001rest, bellezza1990exact}.
\begin{subequations} \label{exact_model} 
\begin{align}
EI \frac{\partial^4 w(x,t) }{\partial x^4} + \rho \frac{\partial^2 w(x,t)}{\partial{t^2}} + \rho x \ddot{\theta} &= 0 \label{7a}\\
\tau (t) - J \ddot{\theta} &= 0 \label{7b}
\end{align}
\end{subequations}
where $\theta(t)$ is the angle between the x-axis and the axis connecting the origin and center of mass position (in $rad$), $J = \left(J_0 + \rho \frac{l^3}{3} + J_p + m_p l^2\right)$ is the total inertia of the flexible link. The partial differential equation \eqref{7a} satisfies the boundary conditions given in \eqref{boundary conditions}:
\begin{subequations} \label{boundary conditions}
\begin{align}
 w(0,t) &= 0 \\
EI \frac{\partial^2 w(0,t)}{\partial x^2} &= J_0 \left[\ddot{\theta} + \frac{\partial^2}{\partial t^2} \left(\frac{\partial w(0,t)}{\partial x} \right)\right] - \tau(t)\\
EI \frac{\partial^2 w(l,t)}{\partial x^2} &= -J_p \left[\ddot{\theta} + \frac{\partial^2}{\partial t^2} \left(\frac{\partial w(l,t)}{\partial x} \right) \right] \\
EI \frac{\partial^3 w(l,t)}{\partial x^3} &= m_p \left[l\ddot{\theta} + \frac{\partial^2 w(l,t)}{\partial t^2}\right]
\end{align}
\end{subequations}
Using separation of variables, deformation $w(x,t)$ can be written in a manner that helps in decoupling the space and time variables as given in \eqref{lateral deflection}. 
\begin{align}\label{lateral deflection}
w(x,t) =  \sum_{i=1}^{\infty} \phi_i(x) p_i(t)
\end{align}
Where $\phi(x)$ is a function of spatial coordinate, $p(t)$ represents vibratory motion and it is a function of time and the number of mode shapes is denoted by $n (1, 2,...,\infty)$. There will be an infinite number of assumed modes for any flexible link, with one natural frequency associated with each assumed mode.  But it is impossible to consider all the vibration modes in the system modelling; therefore, only a finite number of vibration modes are considered in the system modelling that best describes the system's response. As a result, equation\eqref{lateral deflection} can be rewritten using a finite number of modes, say $n$:
\begin{align}\label{n_mode_flexure}
w(x,t) =  \sum_{i=1}^{n} \phi_i(x) p_i(t)
\end{align}
\par As the dynamic model is designed with only $n$ flexible modes in consideration, the dynamic model of a single-link flexible manipulator will always contain unmodeled uncertainties.
\par Replacing $w(x,t)$ from equation \eqref{n_mode_flexure} in equation \eqref{exact_model} in free evolution ($\tau = 0$ $\implies$ $\ddot{\theta} = 0$) and solving PDE by variable separable method, we have a set of ODE's as:
\begin{align} \label{diff_phi}
\frac{d^4\phi_i(x)}{dx^4} - {\beta_i}^4 \phi_i(x) &= 0 \\ \label{diff_p} 
\frac{d^2p_i(t)}{dt^2} + {\omega_i}^2 p_i(t) &= 0
\end{align}
where ${\omega_i}$ is the natural frequency of vibration of modes (eigenvalue) and $\phi(x)$ is the eigenfunction. $\beta_i$ is the spatial vibration frequency of assumed modes.
\begin{align}
\omega_i^2 = \beta_i^4 \frac{EI}{\rho} 
\end{align}
$\beta_1$, $\beta_2$, $\cdots$, $\beta_n$ are the first $n$ roots of the characteristics equation in \eqref{characterstics_eq}.
\begin{align}
 &(c \hspace{0.1cm} sh - s\hspace{0.1cm} ch) - \frac{2m_p}{\rho}\beta s\hspace{0.1cm} sh  - \frac{2J_p}{\rho}\beta^3c \hspace{0.1cm} ch - \frac{J_0}{\rho}\beta^3(1 + c \hspace{0.1cm} ch) \nonumber \\ 
 &- \frac{m_p}{\rho^2}\beta^4(J_0 + J_p)(c \hspace{0.1cm} sh - s \hspace{0.1cm} ch) + \frac{J_0 J_p}{\rho^2}\beta^6(c \hspace{0.1cm} sh + s \hspace{0.1cm} ch) \nonumber \\ 
 &- \frac{J_0J_pm_p}{\rho^3}\beta^7(1 - c \hspace{0.1cm} ch) = 0   \label{characterstics_eq}
\end{align} 
where, $c = \cos(\beta l)$, $ch = \cosh(\beta l)$, $s = \sin(\beta l)$ and $sh = \sinh(\beta l)$.
\par The infinite-dimensional model in \eqref{exact_model}, is approximated with a finite-dimensional model using the modal analysis discussed above. We have considered only a finite number of flexible modes of vibration for study, as shown in \eqref{n_mode_flexure}. By using the equations \eqref{boundary conditions} and \eqref{n_mode_flexure} we get a generalized finite-dimensional dynamic model of a single-link flexible manipulator as given by \eqref{matrix_form}. 
\begin{align}
M\ddot{q}(t) + D\dot{q}(t) + Kq(t) = \Bar{B}\tau(t) \label{matrix_form}
\end{align}
where, $q(t) = (\theta, \hspace{1pt}p_1, ....\hspace{1pt} p_n)^T \in \mathbb{R}^{n+1}$, $M \in \mathbb{R}^{(n+1)\times (n+1)}$, $D \in \mathbb{R}^{(n+1)\times (n+1)}$, $K \in \mathbb{R}^{(n+1)\times (n+1)}$, $\Bar{B} \in \mathbb{R}^{(n+1)\times 1}$, and $\tau(t) \in \mathbb{R}$.
\begin{align*}
M =  \begin{bmatrix}
J & 0 \\
0 &I
\end{bmatrix}, \hspace{0.08cm} D =  \begin{bmatrix}
0 & 0 \\
0 & 2\zeta \Omega
\end{bmatrix}, \hspace{0.08cm}
K = \begin{bmatrix}
0 & 0\\
0 & \Omega^2
\end{bmatrix}, \hspace{0.08cm} \Bar{B}= \begin{bmatrix}
1 \\
\phi'(0)
\end{bmatrix}
\end{align*}
$\Omega = \text{diag}\{\omega_1, .....,\omega_n\}$, 
$\phi'(0) = ({\phi'}_1(0),...\phi'_n(0))^T$.
Where $\phi_i'(0) = \frac{\partial \phi_i(x)}{\partial x}$ at $x = 0$ and $i = 1,2,....,n$ denotes the assumed modes, and $\zeta$ is the damping coefficient.
\par The measured output of the single-link flexible manipulator are the clamped joint angle $\theta_c(t)$ (in $rad$) and tip angle $\theta_t(t)$ (in $rad$) which can be expressed using the states of the dynamic model as:
\begin{align}
\theta_c(t) &= \theta(t) + \sum_{i=1}^{n} \phi_i'(0) p_i(t) \label{clamped_position} \\
\theta_t(t) &= \theta(t) + \sum_{i=1}^{n} \frac{\phi_i(l)}{l} p_i(t) \label{tip_position}   
\end{align} 
\\Equation \eqref{matrix_form} can be transformed to the state space model as:
\begin{align}
\dot{x}(t) &= Ax(t) + Bu(t) \label{state_space_model} \\
y(t) &= Cx(t) \label{output}
\end{align}
where, $x(t) \in \mathbb{R}^{(2n+2)}$, $A \in \mathbb{R}^{(2n+2)\times (2n+2)}$, $B \in \mathbb{R}^{(2n+2)\times 1}$, $C \in \mathbb{R}^{2 \times(2n+2)}$, $y(t) \in \mathbb{R}^{2}$ denotes the output of the system, and $u(t) \in \mathbb{R}$ represents the input to the system.
\begin{align*}
x(t) = \begin{bmatrix}
    \vartheta(t) \\
    \dot{\vartheta}(t)
\end{bmatrix}, \hspace{0.5cm} y(t) = \begin{bmatrix}
    \theta_c(t) \\
    \theta_t(t)
\end{bmatrix}
\end{align*}
Where, $\vartheta(t)= [\theta(t),p_1(t),p_2(t),\ldots,p_n(t)]^T$.
\begin{align}
 A &= \begin{bmatrix}
0 & I \\
-M^{-1}K & -M^{-1}D
\end{bmatrix}, \hspace{0.5cm}
B = \begin{bmatrix}
0\\
M^{-1}\Bar{B}
\end{bmatrix}   \label{system_matrix} \\ 
C &= \begin{bmatrix}
1 & \phi'_1(0) & \phi'_2(0) & \ldots & \phi'_n(0) & 0 & 0 & \ldots\\
1 & \frac{\phi_1(l)}{l} & \frac{\phi_2(l)}{l} & \ldots & \frac{\phi_n(l)}{l} & 0 & 0 & \ldots
\end{bmatrix} \label{output_matrix} \\
u(t) &= \tau(t) \nonumber
\end{align}
\section{Composite Control Design}
\label{sec:composite_control}
\subsection{Sliding Mode Control Design} 
In this section, a sliding mode control law is designed to control the position of the single-link flexible manipulator. The sliding function is chosen as given in equation \eqref{sliding_surface}.
\begin{align}
\sigma(t) =\Gamma \left[x(t) - x_d(t)\right] \label{sliding_surface}
\end{align}
where, $\Gamma > 0 (\in \mathbb{R}^{1\times (2n+2)})$ is a constant, which is to be designed such that the system becomes stable when confined to $\sigma(t) = 0 $, $x_d(t)$ is the desired position of states. 
\par Taking the time derivative of $\sigma(t)$ in \eqref{sliding_surface} and using \eqref{state_space_model}:
\begin{align}
\dot{\sigma}(t) &= \Gamma \left[\dot{x}(t) - \dot{x_d}(t)\right]   \label{sigma_dot} \\
&=  \Gamma \left[Ax(t) + Bu(t) - \dot{x_d}(t)\right] \nonumber 
\end{align}
The proposed control law $u(t)$ has two components, nominal control $u_{nom}$ and discontinuous control $u_{disc}$. The expression for $u(t)$ is given in equation \eqref{control_input}.
\begin{multline}
u(t) =  \underbrace{(\Gamma B)^{-1}\left[-\Gamma A x(t) + \Gamma \dot{x_d}(t)\right] }_{u_{\text{nom}}} \\ - \underbrace{(\Gamma B)^{-1} \left[ k_1\sigma(t) + k_2 \text{sgn}(\sigma(t) ) \right]}_{u_{\text{disc}}} \label{control_input}
\end{multline}
Where, $k_1 \hspace{1pt}\text{and} \hspace{1pt} k_2 > 0(\in \mathbb{R})$ are constants to be designed.
\begin{lemma}[Finite-time lemma \cite{yu2005continuous}] \label{lemma1}  
Considering a continuous time system $\dot{\Psi} = f(\Psi)$, $\Psi \in \mathbb{R}^n$ with zero as the equilibrium point. Let us choose a positive definite Lyapunov candidate function $\mathcal{V}(\Psi): \mathbb{R}^n \rightarrow \mathbb{R}$, with $\alpha_2>0$, $\chi \in (0,1)$, and an open vicinity of origin $\Delta_0 \subseteq \mathbb{R}^n$, such that the inequality in \eqref{FT_lemma_1} is satisfied.
\begin{align}
\dot{\mathcal{V}}(\Psi)\leq - \alpha_1 \mathcal{V}(\Psi) - \alpha_2 \mathcal{V}^{\chi}(\Psi)  ; \quad \Psi \in \Delta_0 \ \{0\},  \label{FT_lemma_1}
\end{align}
then we can say that the equilibrium point is finite-time stable. Further, if $\Delta_0 = \mathbb{R}^n$ then the global finite-time stability of the equilibrium point is guaranteed.
\end{lemma}
\begin{theorem} \label{theorem_stability}
Consider the state space model in \eqref{state_space_model} and the sliding function in \eqref{sliding_surface}. Under the influence of the presented controller, \eqref{control_input}, the sliding phase will be attained in finite time (i.e., $\sigma(t) = 0$), and the system states will converge asymptotically to the desired position.
\end{theorem} 
\begin{proof}
Let us define a Lyapunov function $V_1$ as:
\begin{align}
V_{1}(t) = \frac{1}{2}\sigma^2(t) .  \label{V1}    
\end{align}
The time derivative of $V_{1}(t) $ gives
\begin{align}
\dot V_{1}(t) = \sigma(t) \dot \sigma(t) .  \label{V1d_1}   
\end{align}
From equation \eqref{sigma_dot} put $\dot \sigma(t) $ in \eqref{V1d_1} and using \eqref{state_space_model}:
\begin{align}
\dot V_{1}(t) &= \sigma(t) \Gamma\left ( Ax(t) + Bu(t) - \dot{x}_d(t) \right) .  \label{V1d_2}
\end{align}
Putting $u(t)$ from \eqref{control_input} into \eqref{V1d_2}:
\begin{align} \nonumber
\dot V_{1}(t) &= \sigma(t) \left(- k_1\sigma(t) - k_2 \text{sgn}(\sigma(t) ) \right)   \\ \nonumber
&= -k_1\sigma^2(t) - k_2|\sigma(t)| \\ \nonumber
&= -2k_1\frac{\sigma^2(t)}{2} - 2k_2\frac{\left(|\sigma^2(t)|\right)}{2} \\ 
\dot V_{1}(t) &= -\alpha_1 V_1(t) - \alpha_2 V_1^{\frac{1}{2}}(t) \label{V1d_3}
\end{align}
where $\alpha_1 = 2k_1$, $\alpha_2 = 2k_2$ and $\chi = 1/2$. From equation \eqref{V1d_3}  it is clearly visible that it satisfies lemma \ref{lemma1}'s finite time inequality equation. Thus, it can be inferred that the sliding variable in equation \eqref{sliding_surface} converges to zero in finite time, thereby guaranteeing the convergence of system state $x(t)$ to the desired position $x_d(t)$. 
\end{proof}
The control input in equation \eqref{control_input} can be equivalently written as:
\begin{multline}
  u(t) =  -(\Gamma B)^{-1}\left[\Gamma A + k_1\Gamma \right]x(t) \\
  - (\Gamma B)^{-1} \left[k_2 \text{sgn}(\Gamma x(t) - \Gamma x_d(t) ) \right] \\
  + (\Gamma B)^{-1} \left[\Gamma \dot{x}_d(t) + k_1\Gamma x_d(t) \right]\label{control_input_expanded}  
\end{multline}
The sliding mode control law in \eqref{control_input_expanded} needs the system states for closed-loop design. But the system under consideration does not have all the required states available for the measurement. Therefore, an observer is to be designed to estimate the states required to make a closed-loop control law implementable. As the control input in \eqref{control_input_expanded} needs estimation of some linear function of states, therefore instead of designing a state observer, a linear state function observer is proposed such that the output of the functional observer can be directly used in the controller.
\subsection{Functional Observer}
\par This section introduces the functional observer, a linear state function observer that estimates the linear combination of states required by the control input function. 
\par It is required to make an estimate of the linear combination of state $Fx(t)$, which is expressed using $g(t)(= Fx(t))$. Now, define $F$ using the control input given in \eqref{control_input_expanded} as:
\begin{align*}
F_1 &= -(\Gamma B)^{-1}\left[\Gamma A + k_1\Gamma \right] \\
F_2 &= \Gamma
\end{align*}
Where $F \in \mathbb{R}^{2\times (2n+2)}$ and $g(t) \in \mathbb{R}^{2}$. Hence, $g(t)$ can be expressed  as:
\begin{align}
g(t) = \begin{bmatrix}
F_1 \\
F_2
\end{bmatrix} x(t) \label{estimation_vector}
\end{align}
In order to achieve this linear state function estimation, an observer of the form \eqref{observer} needs to be designed.
\begin{subequations} \label{observer}
\begin{align}
\dot{\hat{\eta}}(t) &= N\hat{\eta}(t) + Ly(t) + Hu(t) \label{obsever_state_estimate} \\
\hat{g}(t) &= Gy(t) + D\hat{\eta}(t) \label{observer_output}
\end{align}    
\end{subequations}
where, $\hat{\eta}(t) \in \mathbb{R}^{v}$ is a state vector. $\hat{g}(t) \in \mathbb{R}^{2}$ is the desired estimate of functional. $N \in \mathbb{R}^{v\times v}$, $L \in \mathbb{R}^{v\times 2}$, $H \in \mathbb{R}^{v}$, $G \in \mathbb{R}^{2\times 2}$, and $D \in \mathbb{R}^{2\times v}$ are unknown matrices. 
\\The output $\hat{g}(t)$ of \eqref{observer_output} is said to estimate $Fx(t)$ in an asymptotic manner if
\begin{align}
 \lim_{t\to\infty} [\hat{g}(t) - Fx(t)] = 0 \label{definition}
\end{align}
Now let us suppose that if $\hat{\eta}(t)$ estimates the linear function of $x(t)$ as $\eta(t) = Tx(t)$ (where $T \in \mathbb{R}^{v\times (2n+2)} $) then, $\hat{g}(t)$ estimates the $Fx(t)$  for which we have the theorem \ref{Observer_theorem}.
\begin{theorem}\label{Observer_theorem}
The completely observable $v^{th}$ order observer will estimate $g(t) = Fx(t)$ if and only if the following conditions are satisfied:
\begin{enumerate}
\item $N$ must be a Hurwitz matrix
\item $TA - NT - LC = 0$
\item $H = TB$
\item $F = GC + DT$
\item $v \geq rank(F - GC)$
\end{enumerate}
where $F \in \mathbb{R}^{2 \times (2n+2)}$ is the linear state function gain matrix and $T \in \mathbb{R}^{v \times (2n+2)}$ is the unknown matrix which is to be determined.
\end{theorem}
\begin{proof}
\cite{darouach2000existence}
\end{proof} 
\subsection{Proposed Functional Observer-based Sliding Mode Control}
This section proposes a composite control law using the sliding mode design and functional observer output.
The error between the linear function estimates is expressed as e(t), as given in \eqref{error_of_estimates}.
\begin{align}\nonumber
e(t) &= \eta(t) - \hat{\eta}(t)\\ 
e(t) &= Tx(t) - \hat{\eta}(t) \label{error_of_estimates}
\end{align}
Using equations \eqref{state_space_model}, \eqref{obsever_state_estimate} in the derivative of $e(t)$ in \eqref{error_of_estimates}, we get:
\begin{align}\nonumber
\dot{e}(t) &= T\dot{x}(t) - \dot{\hat{\eta}}(t) \\
\dot{e}(t) &= T(Ax(t) + Bu(t)) -  (N\hat{\eta}(t) + Ly(t) + Hu(t))  \label{derivative_error}
\end{align}
On simplifying equation \eqref{derivative_error} we get:
\begin{align}
\dot{e}(t) &= (TA - NT - LC)x(t) + Ne(t) \label{error_dynamics_simplified}
\end{align}
By substituting $TA - NT - LC = 0$ from theorem \ref{Observer_theorem} in \eqref{error_dynamics_simplified}, we get:
\begin{align}
\dot{e}(t) = Ne(t) \label{error_dynamics}
\end{align}
Using the results in theorem \ref{Observer_theorem} control input $u(t)$ can be rewritten as:
\begin{multline}
u(t) =  \left[1 \hspace{0.25cm} 0\right]Fx(t) - \left[1 \hspace{0.25cm} 0\right]De(t) \\ -\underbrace{(\Gamma B)^{-1}\left[k_2 \text{sgn}(\Gamma x(t) - \Gamma x_d(t)\right] }_{u_{\text{switching control}}}  + \\ \underbrace{(\Gamma B)^{-1} \left[\Gamma \dot{x}_d(t) + k_1\Gamma x_d(t) \right]}_{u_{\text{constant}}} \label{error_dynamics_control_input}    
\end{multline}
Now equation \eqref{state_space_model} is rewritten using \eqref{error_dynamics_control_input}.
\begin{multline}
\dot{x}(t) = Ax(t) + B\left[1 \hspace{0.25cm} 0\right] Fx(t) - \left[1 \hspace{0.25cm} 0\right]De(t) \\  
+ B\underbrace{\left[u_{\text{switching control}} - u_{\text{constant}}\right]}_{u_{\text{bounded}}}  \label{state_space_with_error}
\end{multline}
A composite system is formed using \eqref{error_dynamics} and \eqref{state_space_with_error}.
\begin{multline}
\begin{bmatrix}
\dot{x}(t) \\
\dot{e}(t)
\end{bmatrix} = \underbrace{\begin{bmatrix}
A + B\left[1 \hspace{0.20cm} 0\right] F & - B\left[1 \hspace{0.20cm} 0\right]D\\
0  & N
\end{bmatrix}}_{A_{C}} \begin{bmatrix}
x(t)\\
e(t)
\end{bmatrix} \\
+ \underbrace{\begin{bmatrix}
    B \\
    0
\end{bmatrix}}_{B_{C}}u_{bounded} \label{composite_system}
\end{multline}
Where, $A_C\in\mathbb{R}^{(2n+2+v) \times (2n+2+v)}$, and $B_C \in \mathbb{R}^{(2n+2+v)}$.
\\ If observer matrix $N$ and system matrix $A$ have distinct eigenvalues, then $TA-NT-LC= 0$ will have a solution for $T$. Also, if the composite system matrix $A_C$ has all the eigenvalues in the plane's left half, the system will be uniformly ultimate bounded. Hence, the observer matrix $N$ is chosen so the composite system matrix has stable eigenvalues.
\\By using the theorem \ref{Observer_theorem} and the condition of stable eigenvalues for the composite system matrix $A_C$ in \eqref{composite_system} the observer matrices can be obtained. Hence, the control input $u(t)$ can be further rewritten using the observer output obtained in \eqref{observer_output}.
\begin{multline}
  u(t) =  \begin{bmatrix}
      1 & 0
  \end{bmatrix}\hat{g}(t) - (\Gamma B)^{-1} \left[k_2 \text{sgn}(\begin{bmatrix}
      0 & 1
  \end{bmatrix}\hat{g}(t) - \Gamma x_d(t) ) \right] \\
  + (\Gamma B)^{-1} \left[\Gamma \dot{x}_d(t) + k_1\Gamma x_d(t) \right]\label{control_input_using_estimation}  
\end{multline}
\par The state space model in \eqref{state_space_model} is of $(2n+2)$ order, designing the control for a large value of $n$ results in a complex and difficult-to-implement control law. Therefore, in this paper, the proposed control in \eqref{control_input_using_estimation} is designed by considering only the first two assumed modes $(n=2)$, and hence the system order considered for designing the proposed control is of \textit{sixth order}. 
\par The proposed control law in \eqref{control_input_using_estimation} designed using the system having two assumed modes, is tested for the system with a larger value of n i.e. considering the dynamic model with more number of modes.

\section{Simulation and Results}
\label{sec:simulations}
This section includes the numerical simulations and results that demonstrate the effectiveness of the presented control approach for the single-link flexible manipulator. This paper simulates the developed control law for the first five vibrational modes. The state space representation for the first five assumed modes obtained by considering $n=5$ in \eqref{state_space_model} is given in \eqref{5_mode_state_space_model}.
\\ Where, $X(t) \in \mathbb{R}^{12}$, $\Tilde{A} \in \mathbb{R}^{12\times 12}$, $\Tilde{B} \in \mathbb{R}^{12\times 1}$, $\Tilde{C} \in \mathbb{R}^{2 \times 12}$, system output is denoted by $y(t) \in \mathbb{R}^{2 \times 1}$ and $u(t) \in \mathbb{R}$ represents the control input to the system.
\begin{align}
\dot{X}(t) &= \Tilde{A}X(t) + \Tilde{B}u(t) \label{5_mode_state_space_model} \\
y(t) &= \Tilde{C}X(t) \label{5_mode_output}
\end{align}
\begin{align*}
X(t)&=[\theta(t),p_1(t),\ldots,p_5(t),\dot{\theta}(t),\dot{p}_1(t),\ldots,\dot{p}_5(t)]^T  \\
y(t) &= [\theta_c(t), \theta_t(t)]^T
\end{align*}
The proposed control law in \eqref{control_input_using_estimation} designed for $n=2$, is applied to the system in \eqref{5_mode_state_space_model}. The physical parameter specifications of the single-link flexible manipulator are given in table \ref{flexible_link_parameter}.
\begin{table}[h]
\centering
\caption{Physical Parameters of single-link flexible manipulator}
\label{flexible_link_parameter}       
\begin{tabular}{cc | cc | cc}
\hline\noalign{\smallskip}
Parameters & Value  & Parameters &  Value & Parameters &  Value\\
\noalign{\smallskip}\hline\noalign{\smallskip}
$\rho$ & 0.5 & $\omega_2$ & 55.88  & $\phi'_4(0)$ & 3.8529 \\
l & 1   & $\omega_3$ & 101.36  & $\phi'_5(0)$ & 2.4422 \\
$m_p$ & 0   & $\omega_4$ & 177.66  & $\phi_1(l)$ & 0.3214 \\
$J_p$ & 0  & $\omega_5$ & 286.84 & $\phi_2(l)$ & -1.6407 \\
$J_0$ & 0.002 & $\phi'_1(0)$ & 32.8184   & $\phi_3(l)$ & 2.4586 \\
$EI$ & 1   & $\phi'_2(0)$ & 10.4096 & $\phi_4(l)$ & -2.3010 \\
$\omega_1$ & 20.53  & $\phi'_3(0)$ & 6.1588 & $\phi_5(l)$ & 2.1568  \\
$\zeta$ & 0.05 &  &  &  & \\
\noalign{\smallskip}\hline
\end{tabular}
\end{table}
\\ The observer designed for the state space model in \eqref{state_space_model} and \eqref{output} by considering $n=2$ has order $2$. The observer matrices for an $2^{nd}$ order functional observer are chosen as:
\begin{align*}
G &= \begin{bmatrix}
 216.8704 & -10.7626 \\
    0.1858 &   0.0924   
\end{bmatrix}, \hspace{0.05cm} N = \begin{bmatrix}
    -0.5  &  2 \\
   -2  & -0.5
\end{bmatrix},
\\
L &= \begin{bmatrix}
    1 & 0\\
    0 & 1 
\end{bmatrix}, \hspace{0.05cm} D = \begin{bmatrix}
    -984.9503 & 159.5181 \\
    9.6574  & -1.0438
\end{bmatrix}, \hspace{0.05cm} H = \begin{bmatrix}
    0.5678 \\
   -0.7321
   \end{bmatrix}
\end{align*}
For the chosen observer matrices the composite matrix $A_C$ has all its eigenvalues in the left-half plane, which guarantees the stability of the composite system.
\begin{table*}[h]
\begin{align*}
F &= \begin{bmatrix}
-429.6914 & -149.5454 & -829.1284 & -62.5493 & 2.3555 & 6.8609 \\
6.3461 & 1.8134 & 4.1048 & 0.8301 & -0.0635 & -0.1765
\end{bmatrix}  \\
T &= \begin{bmatrix}
0.5882 & -0.1581 & -0.0039 & 0.0969 & -0.0004 & 0.0005 \\
-0.3529 & -0.1216 & -0.0181 & 0.3183 & -0.0788 & -0.0033
\end{bmatrix} \\
X(0) &= \setcounter{MaxMatrixCols}{12} \begin{bmatrix}
\pi/8 & 0.001 & 0.002 & 0.002 & 0.002 & 0.001 & 0 & 0.0001 & 0.0002 & 0.0002 & 0.0003 & 0.0002
\end{bmatrix}^T 
\end{align*}
\end{table*}
\par The composite control design parameters are presented in table \ref{control_parameters}.
\begin{table}[h]
\centering
\caption{Control Parameters for Composite Control Design}
\label{control_parameters}       
\begin{tabular}{cc}
\hline\noalign{\smallskip}
Parameters & Value \\
\noalign{\smallskip}\hline\noalign{\smallskip}
$\Gamma$ & $\left[
6.3461\hspace{0.15cm}1.8134\hspace{0.2cm}4.1048\hspace{0.15cm}0.8301\hspace{0.1cm}-0.0635\hspace{0.1cm}-0.1765
\right]$ \\
$k_1$ & 67.71 \\
$k_2$ & 0.001 \\
\noalign{\smallskip}\hline
\end{tabular}
\end{table}
\\ Initial state trajectory conditions are denoted as $X(0)$.
\par The simulation is performed for the system in \eqref{5_mode_state_space_model} and the proposed control input in \eqref{control_input_using_estimation} for $n=2$ is applied to it. The simulation is being performed for both regulation and tracking problems.
\subsection{Regulation Problem}
\begin{figure}[h]
\centering
\includegraphics[width=8.8cm,height=8.9cm]{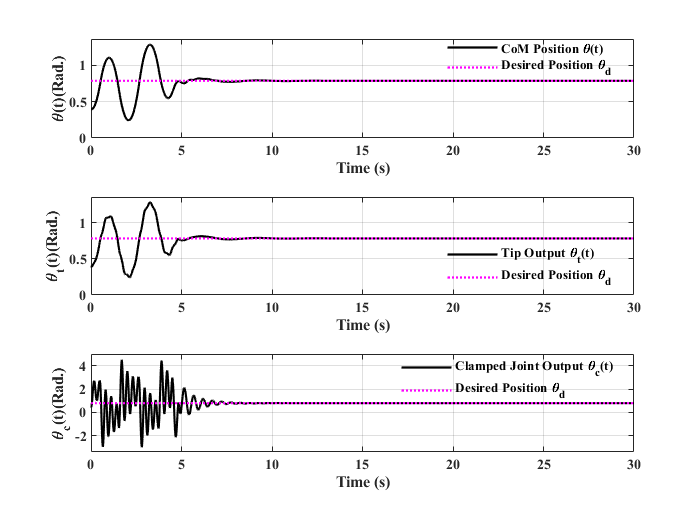}
\caption{Position Angles for Regulation Problem}
\label{fig:regulation_position}
\end{figure}

The reference values for the angle are chosen as:
\begin{align*}
\theta_{d} = \frac{\pi}{4} \hspace{0.25cm} \text{rad.}
\end{align*}
Figure \ref{fig:regulation_position} shows the convergence of tip position $\theta_t(t)$ to the desired position $\theta_{d}$ with vibrations suppressed. The figure also shows the plots for clamped joint angle $\theta_c(t)$ and center of mass position $\theta(t)$.
\par
\begin{figure}[h]
\centering
\includegraphics[width=8.5cm,height=4.5cm]{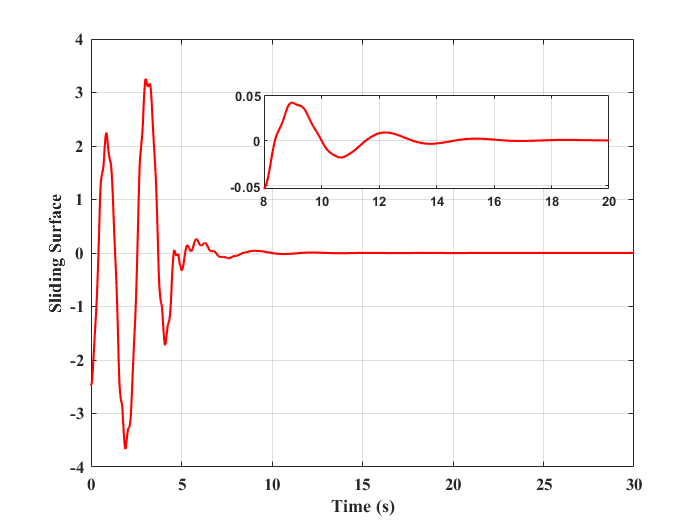}
\caption{Sliding Surface for Regulation Problem}
\label{fig:regulation_sliding}
\end{figure}

 The plot for the sliding variable versus time is shown in figure \ref{fig:regulation_sliding}. Figure indicates that the sliding variable converges to zero in finite time. 
\begin{figure}[h]
\centering
\includegraphics[width=8.5cm,height=4.5cm]{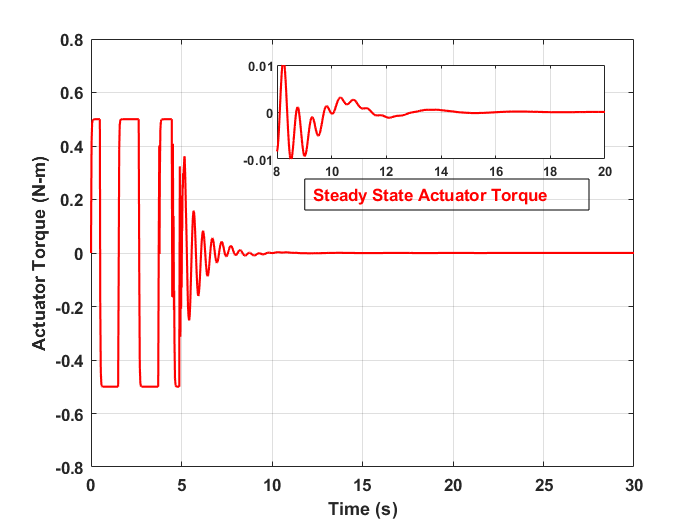}
\caption{Control Input for Regulation Problem}
\label{fig:regulation_control}
\end{figure}
\\ Figure \ref{fig:regulation_control} shows the actuator torque applied the manipulator. The actuator torque applied is well within the bound of $\pm \hspace{0.1cm}0.5 N-m$ i.e. the applied control input is bounded.
\subsection{Tracking Problem}
\begin{figure}[h]
\centering
\includegraphics[width=8.8cm,height=8.9cm]{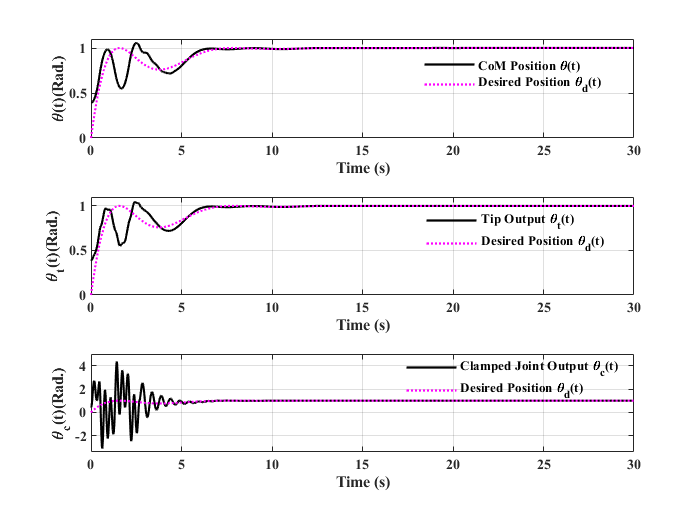}
\caption{Position Angles for Tracking Problem}
\label{fig:Tracking_position}
\end{figure}

\begin{figure}[h!]
\centering
\includegraphics[width=8.5cm,height=4.5cm]{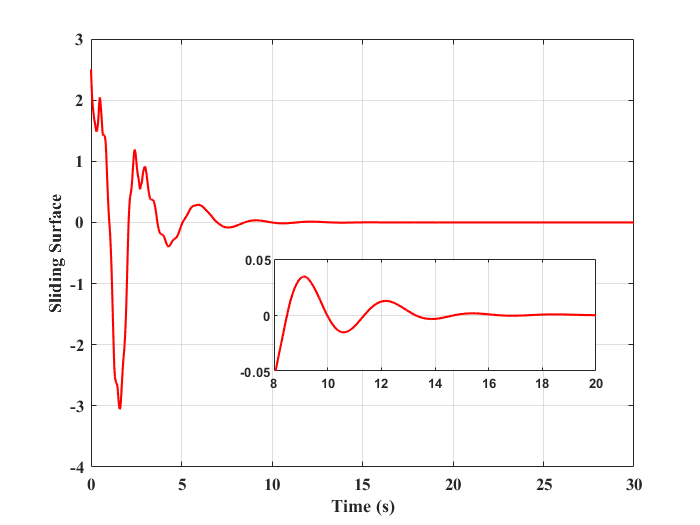}
\caption{Sliding Surface for Tracking Problem}
\label{fig:Tracking_sliding}
\end{figure}

The desired trajectory for the position of a manipulator is chosen as:
\begin{align*}
\theta_{d}(t) = e^{-0.5t}\sin(t)+(1-e^{-0.5t}) \hspace{0.25cm} \text{rad.}
\end{align*}
Figure \ref{fig:Tracking_position} shows the convergence of tip position $\theta_t(t)$ to the desired trajectory $\theta_{d}(t)$ with vibrations being suppressed. The figure also shows the trajectories for clamped joint angle $\theta_c(t)$ and center of mass position $\theta(t)$.
\\The plot of the sliding variable vs time is shown in figure \ref{fig:Tracking_sliding}. It is evident from the figure that the sliding variable converges to zero in a finite amount of time.
\begin{figure}[h!]
\centering
\includegraphics[width=8.5cm,height=4.5cm]{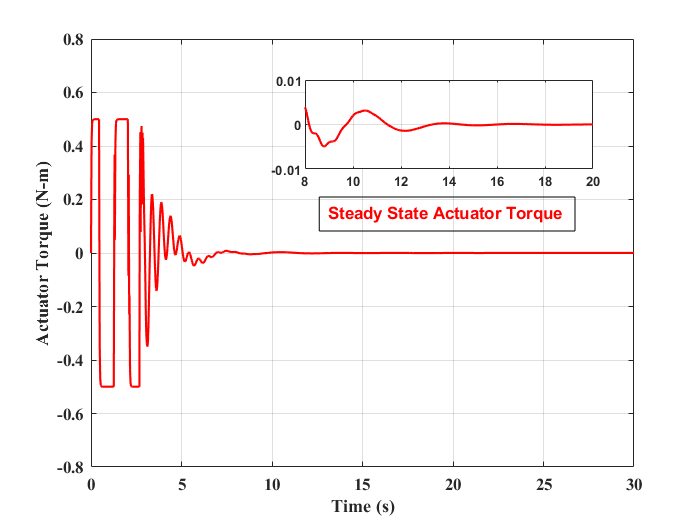}
\caption{Control Input for Tracking Problem}
\label{fig:Tracking_control}
\end{figure}
\\ The plot of actuator torque applied to the manipulator with respect to time is shown in figure \ref{fig:Tracking_control}. The figure shows that the actuator torque has a lower limit of 0.5N m and an upper limit of +0.5N m.

\section{Conclusion}\label{sec:conclusion}
This paper proposes a sliding mode approach based on a functional observer for controlling the position of a single-link flexible manipulator. The proposed control is designed using the first two assumed modes, and its effectiveness is tested using numerical simulation for the dynamic model with the first five vibration modes considered in the modelling. The simulation results indicate that the presented control technique efficiently controls the position of the single-link flexible manipulator.

\bibliographystyle{IEEEtran}
\bibliography{references}

\end{document}